\documentclass[onecolumn]{revtex4}
\usepackage{graphicx}
\usepackage{amssymb}
\usepackage{xcolor}

\def\al{\alpha}
\def\be{\beta}
\def\ga{\gamma}

\def\th{\theta}
\def\la{\lambda}
\def\si{\sigma}

\def\la{\lambda}
\def\vp{\varphi}
\def\ve{\varepsilon}
\def\pa{\partial}

\def\bx{{\bf{x}}}
\def\by{{\bf{y}}}
\def\bp{{\bf{p}}}

\def\EE{{\rm E}_{10}}
\def\KE{{{\rm K}(\EE})}

\def\cO{{\mathcal O}}
\def\cQ{{\mathcal Q}}

\def\11{{\mathbb 1}}

\def\ri{{\rm i}}
\def\rd{{\rm d}}

\def\eV{\,\rm eV}

\def\GeV{\,\rm GeV}

\def\MPL{M_{\rm Pl}}

\def\tPL{t_{\rm Pl}}
\def\MBPS{M_{\rm BPS}}

\def\beq{\begin{equation}}
\def\eeq{\end{equation}}
\def\bea{\begin{eqnarray}}
\def\eea{\end{eqnarray}}
\def\nn{\nonumber}

\def\Ra{\Rightarrow}

\def\ri{\text{i}}

\begin{document}
\title{Supermassive gravitinos and giant primordial black holes}
\author{Krzysztof A. Meissner$^1$ and Hermann Nicolai$^2$}
\address{
$^1$Faculty of Physics,
University of Warsaw\\
Pasteura 5, 02-093 Warsaw, Poland\\
$^2$Max-Planck-Institut f\"ur Gravitationsphysik
(Albert-Einstein-Institut)\\
M\"uhlenberg 1, D-14476 Potsdam, Germany\\
}

\vspace{3mm}

\begin{abstract} 
\noindent
We argue that the stable (color singlet) supermassive gravitinos proposed in our previous
work can serve as seeds for giant primordial black holes. These seeds are hypothesized
to start out as tightly bound states of fractionally charged gravitinos in the radiation 
dominated era, whose formation is supported by the universally attractive combination 
of gravitational and electric forces between the gravitinos and anti-gravitinos (reflecting 
their `almost BPS-like' nature).  When lumps of such bound states coalesce and 
undergo gravitational collapse, the resulting mini-black holes can escape Hawking 
evaporation if the radiation temperature exceeds the Hawking temperature.
Subsequently the black holes evolve according to an exact solution of Einstein's equations, 
to emerge as macroscopic black holes in the transition to the matter dominated era,
with masses on the order of the solar mass or larger. The presence of these seeds at 
such an early time provides ample time for further accretion of matter and radiation,
and would imply the existence of black holes of almost any size in the universe, 
up to the observed maximum.
\end{abstract}
\maketitle

\vspace{5mm}

\section{Introduction.} 

The origin of large (galactic) black holes, present already in the early Universe 
has been a long standing puzzle, see {\em e.g.} \cite{GiantBH} for information on 
the most recently discovered behemoth black hole, \cite{R0} for a generally accessible
update and overview, and \cite{R1,R2,R3} and references therein for more recent work.
It seems generally agreed that such large black holes cannot form by the usual stellar 
processes ({\em i.e.} gravitational collapse of stars and subsequent accretion of mass), 
but must have originated from some other source. One possible explanation 
is that black holes were already present from the very beginning of the matter dominated 
period, and in sufficient numbers and with sufficiently large masses to be able to grow 
further by accretion to very large sizes already a few hundred million years after the Big Bang. 
Various mechanisms have been proposed and discussed
towards solving this problem, most of them based on extrapolations of known physics,
such as {\em e.g.} large random density fluctuations in the early universe, see  \cite{CKSY} 
for a comprehensive recent review with many further references. That review also discusses
different observational consequences and constraints, while emphasizing that ``the limits are 
constantly changing as a result of both observational and theoretical developments".
From a more theoretical perspective, a mechanism based on bubble formation during
inflation was recently put forward in \cite{V,HD}, but differs essentially from the one presented 
here, because there the substantive part of black hole growth must take place {\em before}
the onset of the radiation phase. At any rate, the crucial question remains
whether an explanation can be found in terms of known physics, or 
whether an explanation necessarily involves essentially new physics.

In this paper we present a new proposal towards addressing this problem 
which can complement
existing proposals in that it does not rely on random processes, such as density 
fluctuations or bubble formation, but invokes {\em new} physics. It is based
on the conjectured existence of  certain supermassive particles (gravitinos)
that allow for the formation of black holes already during the early radiation phase, well 
before decoupling. There are two necessary prerequisites for a mechanism based on
the `condensation' of superheavy particles to work, namely 
\begin{enumerate}
\item the supermassive particles must be absolutely stable against decay into 
         Standard Model matter; and 
\item they must be subject to sufficiently strong attractive forces to enable them to 
         rapidly cluster in sufficient amounts to undergo gravitational collapse. 
\end{enumerate}
Although ans\"atze towards fundamental physics, in particular Kaluza-Klein theory and
string theory, abound in massive excitations that might serve as candidates
for such a scenario, such excitations usually fail to meet the first requirement
(with decay lifetimes on the order of the Planck time $\tPL$),
which is why they are often assumed to play no prominent role in the cosmology 
of the very early universe. Here we will argue that, by contrast, the superheavy gravitinos 
proposed in our previous work \cite{MN2,MN0} can meet both requirements.
That the requisite particles should be gravitinos, rather than some other particle 
species, is perhaps unusual, so let us first explain the reasons for this claim.

Our proposal has its origin in our earlier attempt to understand the observed 
spin-$\frac12$ fermion content of the Standard Model, with three generations 
of quarks and leptons (including three right-chiral neutrinos). It relies on a unification 
scenario  based on a still hypothetical extension of maximally extended 
$N\!=\!8$ supergravity involving the infinite-dimensional duality symmetries 
$\EE$ and $\KE$ \cite{MN0,MN2,KN}  (this proposal itself has its origins
in much earlier work \cite{GM,NW}).
The enlargement of the known duality symmetries of supergravity and M theory
to the infinite-dimensional symmetries $\EE$ and $\KE$ is absolutely essential here,
because without this extension neither the charge assignments of the quarks and leptons,
nor those of the gravitinos in (\ref{GravCharges}) below could possibly work, 
and stability of the gravitinos against decay could not be achieved. A key feature 
of our proposal, and one that sets it apart from all other unification schemes, 
is that besides the 48 spin-$\frac12$ fermions of the Standard Model, 
the {\em only} other fermions are the eight supermassive gravitinos 
corresponding to the spin-$\frac32$ states of the $N=8$ supermultiplet.
It is thus a {\em prediction}  that the spin-$\frac12$ fermion content of the 
Standard Model will remain unaltered up to the Planck scale -- a prediction that
is (at least so far) supported by the absence of any signs of new physics from LHC,  
and by the fact that the currently known Standard Model couplings 
can be consistently evolved all the way to the Planck scale. Indeed, the detection 
of any new  fundamental spin-$\frac12$ degree of freedom (such as a sterile fourth 
neutrino, or a fourth generation of quarks and leptons, or any of the `{\em -ino}' 
fermions predicted by low energy supersymmetry) would immediately falsify the present scheme.

Evidence for infinite-dimensional duality symmetries of Kac-Moody 
type comes from an earlier BKL-type analysis of cosmological singularities in general relativity
\cite{DH,DHN1}. This has led to the conjecture that M theory in the `near singularity limit'
is governed by the dynamics of an $\EE/\KE$ non-linear $\si$-model \cite{DHN2}.
In this scenario space-time, and with it space-time based quantum field theory and
space-time symmetries would have to be emergent, in the sense that all the relevant
information about space-time physics gets encoded in and `spread over' a 
hugely infinite-dimensional  hyperbolic Kac-Moody algebra.  In particular, this scheme goes 
{\em beyond} supergravity in that the infinite-dimensional $\EE$ duality symmetry replaces,
and quite possibly disposes of, supersymmetry as a guiding principle towards unification.
The fermionic sector of the theory is then governed by the `maximal compact'
(or more correctly, `involutory') subgroup $\KE\subset\EE$, which can be regarded as an 
infinite-dimensional generalization of the usual R-symmetries of extended supergravity 
theories. While an analysis of the bosonic sector of the $\EE/\KE$ model and its dynamics beyond 
the very first few levels is severely hampered by the fact that a full understanding of $\EE$ 
remains out of reach, a remarkable property of its involutory subgroup $\KE$ is the existence
of {\em finite}-dimensional (unfaithful) spinorial representations \cite{DKN,dBHP,KNV}. 
The combined spin-$\frac12$ and spin-$\frac32$ fermionic degrees of freedom 
at any given spatial point are then no longer viewed as fermionic members of the $N\!=\!8$ supermultiplet, but rather as belonging to an (unfaithful) irreducible representation of 
the generalized R-symmetry $\KE$ \cite{DKN,dBHP,KNV}. The link with the physical
fermion states is then made by identifying the known $\KE$ representation with the 
Standard Model fermions at a given spatial point, in the spirit of a BKL-type expansion 
in spatial gradients, as explained for the bosonic sector in \cite{DHN2}.

A crucial feature is now that the gravitinos are predicted 
to  participate in strong and electromagnetic interactions (unlike the sterile gravitinos 
of MSSM-like models with low energy supersymmetry), and that they carry fractional
charges. More precisely,  as a consequence of the group theoretic analysis in 
\cite{MN0,MN2,KN}, the eight massive gravitinos are assigned to the following 
representations of the residual unbroken SU(3)$_c \,\times\,$U(1)$_{em}$  symmetry 
\beq\label{GravCharges}
\left({\bf 3}_c\,,\,\frac13\right) \oplus \left(\bar{\bf 3}_c\,,\,-\frac13\right)
\oplus \left({\bf 1}_c\,,\,\frac23\right) \oplus \left({\bf 1}_c\,,\, -\frac23\right)
\eeq 
These assignments follow from an SU(3)$\,\times\,$U(1) $\subset$ SO(8) decomposition 
of the $N\!=\!8$ supergravity gravitinos, {\em except} for the `spurion' shift of the U(1) charges 
by $\pm\frac16$ that was originally introduced in \cite{GM} for the spin-$\frac12$ members
of the $N\!=\!8$ supermultiplet, in order to make their electric charge assignments agree 
with those of three generations of quarks and leptons (including right-chiral neutrinos). 
As shown in \cite{MN0,MN2,KN}, it is this latter shift which requires enlarging 
the R-symmetry to $K(E_{10})$, and which takes the construction {\em beyond} 
$N\!=\!8$ supergravity and {\em beyond} the confines of space-time based field theory.
All gravitinos are assumed to be superheavy, with masses just below the Planck mass. 
This assumption is plausible because in any scheme avoiding low energy supersymmetry 
and in the absence of grand unification the Planck scale is the natural scale 
for symmetry breaking. Despite their large mass {\em all gravitinos are stable 
against decays into Standard Model matter}, as a consequence of their peculiar 
quantum numbers: there is simply no final state in the  Standard Model into which 
they could possibly decay in compliance with (\ref{GravCharges}) and 
the residual unbroken SU(3)$_c \,\times\,$U(1)$_{em}$ symmetry. This
feature is essentially tied to the replacement of the usual R-symmetry by $\KE$,
because in a standard supergravity context a supermassive gravitino would
not be protected against decay into other particles.

In the present paper we take a more pragmatic approach by simply proceeding
with the assignments (\ref{GravCharges}) as the starting point, but keeping in mind 
that this scheme is strongly motivated by unification and a possible explanation  
of the observed pattern of quark and lepton charge quantum numbers, and thus not 
based on {\em ad hoc} choices. In \cite{MN1,MN3} we have already begun to explore the 
possible astrophysical implications of supermassive gravitinos with the above assignments. 
More specifically, in \cite{MN1} we have proposed the color singlet gravitinos as novel 
dark matter candidates, and discussed possible avenues to search for them 
(in fact, even within the present scenario, this
proposal would hold up, in that the supermassive gravitinos could make up 
a large part, or even all of dark matter, via the black holes into 
which they would have been swallowed). In subsequent work \cite{MN3}  
we showed that the color triplet states in (\ref{GravCharges}) can potentially 
explain the observed ultra-high energy cosmic ray events with energies of up to $10^{21} \eV$ 
via gravitino anti-gravitino annihilation in the crust of neutron stars. In this paper we now turn 
our attention again to the {\em color singlet} gravitinos of charge $\pm\frac23$, to argue that 
they can in addition play a key role in shedding light on the origin of giant black holes in 
the early universe.

The structure of this paper, then is as follows: in section II we show that
quantum mechanically the wave function of a multi-gravitino bound state
is highly unstable against gravitiational collapse. In the following two sections
we study the formation and evolution of mini-black holes during the radiation
era, also deriving numerical estimates. For the evolution we employ a  
generalization of the McVittie solution (on which there is already an ample literature,
see {\em e.g.} \cite{McV0,McV1,McV2,McV3,McV4,McV5,McV6} and references therein).
In the last section we analyze the energy-momentum tensor for this solution,
and show that it has the right form expected for a radiation dominated
universe. We also argue that the `blanket' surrounding the primordial 
black hole can further enhance the growth of massive black holes. 
These last two sections may be of interest in their own right, 
independently of the main line of development of this paper.


\section{Formation of multi-gravitino bound states}

The main new feature of our proposal is that, as a result of the assumed
large mass of the gravitinos, the combined gravitational and electric forces
between any arrangement of gravitinos and anti-gravitinos is {\em universally
attractive}. In natural units we define the BPS-mass $\MBPS$ for the (anti-)gravitino
to be the one for which the electrostatic force between two gravitinos with charges 
$\pm Q_g$ equals their gravitational attraction (modulo sign)
\beq\label{BPS}
Q_g^2  \,=\,  G \MBPS^2   \; ;
\eeq  
we refer to $\MBPS$ as the `BPS-mass' because it is the one relevant for 
extremal Reissner-Nordstr\"om or Kerr-Newman solutions. This equality is written in units 
where $4\pi\epsilon_0=\mu_0/(4\pi)=c=1$ (here it is worthwhile to recall that 
these units, with the addition of $e=\MBPS=1$, were introduced already in 1881 
by George Stoney, probably the first physicist who seriously 
contemplated quantization of charge \cite{Stoney};  the electron was 
discovered only 16 years later, while Planck units were introduced 18 years later). 
As is well known,  $\MBPS$ is {\em not}  the same as the (reduced) Planck mass $\MPL$, 
but differs from it by a factor of the fine structure constant $\al$ (always with $c=1$ from now on):
\beq
\MBPS^2 \,=\, \frac{Q_g^2}{G} \,=\, \frac{Q_g^2}{\hbar} \cdot \frac{\hbar}{G}
                \,\equiv \,  \al\MPL^2\,.
\eeq
where $\al$ differs from the usual fine structure constant $\al_{em}$ by a factor 
$\frac49$ because of the fractional charge, see below.
We will assume that the gravitino mass lies between these two values, {\em i.e.} 
 \beq\label{Mgrav}
 \MBPS   < M_g < \MPL
 \eeq 
 The first of these inequalities is needed to ensure
 that the force between same charge gravitinos remains attractive; for $M_g < \MBPS$ 
 we would have repulsion [because $(1-\be^2)$ in (\ref{V}) becomes negative], 
 and the proposed mechanism would no longer work. Denoting the usual elementary
 charge by $e$ we can thus write for the gravitino charges
\beq
Q_g = \pm \frac23 e = \pm \be G^{\frac12} M_g 
\eeq
with the `BPS-parameter' $\be$ obeying $0 < \be < \frac23$; we will denote the (fixed)
gravitino mass by $M_g$ throughout this paper, whereas generic black hole masses will 
be designated by the letter $m$, where $m$ can also vary with time. The total force 
between two (anti-)gravitinos is thus determined by the combined electric and gravititional 
charges $(1 \pm \be^2) GM_g^2 > 0$, so that even for like charges the force remains 
attractive because the gravitional attraction overwhelms the electrostatic repulsion 
(reflecting the `almost BPS-like' nature of the gravitinos). In this paper we hypothesize 
that it is this universal attraction that leads to the formation of multi-gravitino bound 
states inside the plasma of the radiation dominated phase, starting from small inhomogeneities
in analogy with cluster formation of galaxies. The main difference with the latter
is that, prior to gravitational collapse, we are here initially dealing 
with a {\em quantum mechanical} bound state, not one
that can be understood in terms of Newtonian physics. For two gravitinos the bound state would
be somewhat analogous to positronium, however with the crucial difference that `gravitinium'
can be a longer lived state because the annihilation cross section between two oppositely 
charged (color singlet) gravitinos is very small, of the order 
$\sim M_g^{-2}$ (as follows from inspection of the 
standard tree level Feynman diagram for annihilation into, say, a pair of gravitons,
with one intermediate gravitino  propagator). Note that in principle
positronium can also be long lived, provided the bound state is formed in a state 
of very large radial quantum number \cite{PM} (see (\ref{relax}) below). 

We wish to study the formation of bound states of gravitinos during the 
radiation era in the very early universe. For a proper analysis, and as a first step,
we would now have to go through a first quantized analysis of the massive 
Rarita-Schwinger equation in such a homogeneously and isotropically expanding
background. This task is substantially simplified by our main assumption (\ref{Mgrav})
which allows us to resort to the non-relativistic limit, and by the fact that this 
inequality also implies
\beq\label{MH}
M_g \,>\, H(t)
\eeq
for the Hubble parameter during the radiation era, whence we can also drop the 
usual friction term $\propto H(t) = \dot a(t)/a(t)$ that would normally have to be 
included in the equation of motion. It is therefore enough to consider the free 
Rarita-Schwinger equation for a massive spin-$\frac32$ complex vector spinor, 
which reads
\beq\label{RS}
   i\ga^{\mu\nu\rho} \pa_\nu \psi_\rho + M_g\ga^{\mu\nu} \psi_\nu \,=\, 0
\eeq
From this one immediately deduces the Dirac and constraint equations
\beq
(i\ga^\la \pa_\la -M_g) \psi_\mu = 0 \quad \mbox{and} \quad
\ga^\mu \psi_\mu \,=\, \pa^\mu \psi_\mu \,=\, 0
\eeq
(see {\em e.g.} \cite{dWF} for a more complete account). 
The latter two equations imply a halving of the available degrees of freedom,
and tell us that the vector spinor $\psi_\mu$ carries altogether four helicity degrees 
of freedom, with labels $\si,\tau \in \left\{\pm \frac12,\pm \frac32\right\}$ for both gravitino 
and anti-gravitino. The relevant expansion reads
\bea
\psi_\mu (x) \,&=&\,  \int \frac{\rd^3  \bp}{(2\pi)^{3/2} \sqrt{2E(\bp)}} 
\Big[ e^{ipx} f_\mu^+(p) u_+(p) + e^{ipx} f_\mu^-(p) u_-(p)  \,+\,  \nn\\[2mm]
&& \hspace{3.5cm} + \; e^{-ipx} g_\mu^+(p) v_+(p) \,+\,  e^{-ipx} g_\mu^-(p) v_-(p) \Big] 
\eea
where, of course, $p^2 +M_g^2=0$, and $u_\pm(p)$ and $v_\pm(p)$ are the two positive 
and negative energy solutions of the Dirac equation. The last constraint equations is 
solved by
\beq
f_\mu^\pm(p) = \sum_{\ri} b_{\ri}^\pm (p) \ve^{\ri}_\mu(p) \quad  , \quad
g_\mu^\pm(p) = \sum_{\ri} d_{\ri}^\pm(p) \ve^{\ri}_\mu(p) 
\eeq
with the three linearly independent polarization vectors $\ve^{\ri}_\mu(p)$
satisfying  $p^\mu \ve^{\ri}_\mu(p) = 0$. For the other constraint equation 
we need to impose
\bea
\sum_{\ri} \ga^\mu \ve^{\ri}_\mu(p) \Big[ b_{\ri}^+(p)   u_+(p) \,+\, 
  b_{\ri}^-(p) u_-(p)\Big]  \,&\stackrel{!}{=}& \, 0       
\nn\\[2mm]
\sum_{\ri} \ga^\mu \ve_\mu^{\ri}(p) \Big[d_{\ri}^+(p) v_+(p) \,+\, 
                        d_{\ri}^-(p) v_-(p)\Big]  \,&\stackrel{!}{=}& \, 0   
\eea
thus eliminating four out of the 12 free coefficients $b_{\ri}^\pm(p)$ and $d_{\ri}^\pm(p)$, 
respectively, leaving us with four helicity wave functions for gravitino and anti-gravitino each.
As the spin interactions are not relevant for our approximation there is no need here to be 
any more specific about the parametrization of the helicity wave functions. However, each 
gravitino degree of freedom is exposed to the gravitational and electric background 
generated by the other gravitinos (as well as the surrounding plasma which we can neglect).
In order to incorporate these interactions in lowest order, one performs the standard 
Foldy-Wouthuysen transformation on each component of $\psi_\mu$, which yields a 
non-relativistic one-particle Hamiltonian for each gravitino component.

The corresponding multi-particle Schr\"odinger Hamiltonian therefore reads
\beq\label{H}
H = - \frac{\hbar^2}{2M_g}\sum_i \big( \triangle_{\bx_i} + \triangle_{\by_i}\big) 
     \,+\, V(\bx,\by)
\eeq
with the universally attractive potential (for $\be^2 < 1$)
\beq\label{V}
V(\bx,\by) = - (1-\be^2) \left( \sum_{i\neq j} \frac{GM_g^2}{|\bx_i - \bx_j|} \,+\,
                    \sum_{i\neq j} \frac{GM_g^2}{|\by_i - \by_j|}\right) \,-\,
                   (1+\be^2) \sum_{i,j} \frac{GM_g^2}{|\bx_i - \by_j|}
\eeq
where the positions of the gravitinos and anti-gravitinos are designated
by $\bx_i$ and $\by_j$, respectively. This Hamiltonian acts on a fermionic wave function
$\Psi(\bx_1,\si_1,\dots,\bx_n,\si_p;\by_1,\tau_1,\dots,\by_q,\tau_q)$ which is
is antisymmetric under simultaneous interchange of the position and spin labels
of the gravitinos and anti-gravitinos, respectively. In writing this Hamiltonian 
we have also neglected the fluctuating external electric and magnetic fields 
in the radiation plasma. Likewise, as we already explained, we ignore subleading
spin-orbit and spin-spin interactions that would follow from the Rarita-Schwinger equation
in a fully relativistic treatment (and which would be very complicated).
Finally, we can neglect the effect of the protons and electrons from the 
surrounding plasma (as well as all other Standard Model particles): for them, the 
gravitational interactions are governed by the factors $GM_g m_e\,,\,
GM_g m_p\, ,... \ll GM_g^2$, whence their interactions are completely dominated 
by the purely electromagnetic forces. The latter are, however, screened out because of the overall
electric neutrality of the plasma, and can thus be ignored. 

Evidently the above considerations only apply to superheavy particles
obeying (\ref{Mgrav}) and (\ref{MH}), and would not make any sense at all for ordinary
(Standard Model) particles. For the latter all masses and binding energies are far 
below the temperature of the surrounding plasma, that is $m_e, m_p,... \ll T_{rad}$, 
and also below the Hubble parameter, $m_e, m_p,...\ll H$. 
In that case, the stationary Schr\"odinger equation would 
have to be replaced by a relativistic equation in a time-dependent background, and the 
friction term involving the Hubble parameter $H$  would lead to immediate decay of the 
wave function (as unitarity in the naive sense is violated in a time-dependent background). 

We will not attempt here to investigate in any detail the multi-particle Schr\"odinger 
equation based on  (\ref{H}), which would amount to a quantum analog of the 
computations performed in connection with galaxy structure formation. Nevertheless, 
we can  still make some rigorous statements relying on well known estimates (see {\em e.g.} \cite{LS}).
Namely, it is a rigorous result \cite{LL} that for a system of fermions (that is, particles obeying 
the Pauli principle with a fully antisymmetric wave function) the lowest energy eigenvalue 
of the $N$-particle Hamiltonian 
\beq\label{E0}
E_0(N) := \inf_{|\!|\Psi |\!| =1} \langle \Psi | H | \Psi \rangle
\eeq
(where $N$ is the combined number of gravitinos and anti-gravitinos)
is subject to the upper and lower bounds
\beq\label{E1}
 - AN(N-1)^{\frac43} G^2M_g^5 \hbar^{-2} \,\leq\,  E_0(N) \, \leq \,
       - BN^{\frac13}(N-1)^2 G^2M_g^5 \hbar^{-2}
\eeq
with strictly positive constants $A > B > 0$. Consequently the lowest energy per
particle $E_0(N)/N$ decreases as $\propto - N^{4/3}$ with $N$,
signaling an instability. For a bosonic wave function
the fall-off would be even faster with $E_0(N)/N \propto - N^2$ \cite{LL}. Therefore
the inclusion of spin degrees of freedom (where one combines a partially symmetric 
wave function in the space coordinates with an anti-symmetric wave function in
spin space) cannot improve the situation. The estimate (\ref{E1}) tells us that the
system is unstable, and for sufficiently large $N$ will thus undergo gravitational
collapse, as the fermionic degeneracy pressure is not enough to sustain the system
in a stable equilibrium. Because of (\ref{MH}) the basic instability estimate (\ref{E1}) is not 
affected by the cosmological expansion either.

Now if we consider a bound state of just two gravitinos (a hydrogen-like system) the 
associated `Bohr radius' is only a few orders of magnitude away from the Planck length, to wit
\beq 
a_B \,\sim \, \frac{\hbar^2}{G M_g^3} 
\eeq
which is not too far from the Schwarzschild radius. If the formation of such bound states
took place in vacuum, and the relaxation to the ground state proceeded too fast, 
the resulting mini-black holes  would immediately evaporate by Hawking radiation 
according to the well known formula (see {\em e.g.} \cite{TD})
\beq\label{evap}
t_{evap} \,\sim \, t_{\rm Pl} \left(\frac{m}{\MPL}\right)^3  \,\sim \, 
10^{-42}\, {\rm s} \left(\frac{m}{10^{-9}{\rm kg}}\right)^3 
\eeq
which follows from the Stefan-Boltzmann law upon substitution of the Hawking temperature
\beq\label{BHtemp}
T_{Hawking} = \frac{\hbar}{8\pi Gm}
\eeq
In order to prevent this from happening, and in order to create bigger black 
holes that can survive for longer and start growing, it is therefore necessary for the 
bound states to persist long enough to accrete  a sufficiently large number of 
gravitinos {\em before} gravitational collapse. Meta-stability can be ensured if 
the initial energy of the bound state
is much larger than $E_0(N)$, and consequently its overall extension stays well above 
its Schwarzschild radius for sufficiently long time. Of course, the bound state will eventually 
relax to lower lying bound states by the spontaneous emission of photons and gravitons, 
but this process will take some time. For instance, for positronium 
the average lifetime $\tau$ of a bound state as a function of 
the principal quantum number $n$ scales as (see {\em e.g.} \cite{LL0}, eqs. (7)--(9))
\beq\label{relax}
\tau \sim \, n^4 
\eeq
In comparison with positronium which has a large annihilation cross section, the mutual 
annihilation of (color singlet) gravitinos and anti-gravitinos is further delayed by their 
small annihilation cross section $\sim M_g^{-2}$, which was already 
highlighted above. Extrapolating the above formula 
to the present case thus suggests that, with sufficiently large $n$ at the 
time of formation, we can get lifetimes long enough to bind a large number 
of gravitinos into a meta-stable configuration before the collapse can occur. We also note 
that at this stage (that is, prior to the formation of a black hole)  the absorption 
of protons and electrons from the ambient plasma plays no role, as 
these particles, unlike the gravitinos, will be only very weakly bound.

\section{Collapse of gravitino lumps and mini-black holes}

At this point we have lumps, each corresponding to a quantum mechanical multi-gravitino 
bound state, which are scattered throughout the radiation plasma. Because of the density 
fluctuations and inhomogeneities in the plasma, and as a result of their strong gravitional 
attraction these lumps will eventually coalesce before collapsing into small black holes, a 
microscopic analogue of the clumping of dust into galaxies and stellar matter. In a first
approximation the ensemble of massive lumps can be treated classically ({\em i.e.} 
need not be considered as a single coherent wave function). 
In order to arrive  at a rough estimate of the initial mass of the resulting black holes we 
first estimate the total number of gravitinos contained in a coalesced lump of gravitino
matter. Treating them classically with an average kinetic energy per particle 
equal to the temperature of the plasma we have
\beq\label{Ekin}
\langle E_{kin} (t) \rangle \,\sim \,  N T_{rad}(t)  = N T_{eq} \left(\frac{t_{eq}}{t}\right)^{1/2}
\eeq
where $T_{eq}\sim 1\,$eV and $t_{eq}\sim 40\,000\,{\rm yr} \sim 10^{12}$ s (we find
it convenient to refer all quantities to equilibrium time $t_{eq}$ rather than Planck 
units). The potential energy of $N$ gravitinos and anti-gravitinos is given by
\beq
\langle E_{pot}(t) \rangle \sim - \, N^2 \frac{GM_g^2}{\langle d(t) \rangle} 
\eeq
where for numerical estimates we take $M_g \sim \MBPS$.
The average separation $\langle d(t)\rangle$ between gravitinos and
anti-gravitinos at time $t$ is given by
\beq
\langle d(t) \rangle \sim \left( \frac{M_g}{\rho(t)} \right)^{1/3} \,
\sim \, (10^2 \, {\rm m}) \,  \left(\frac{t}{t_{eq}} \right)^{1/2}
\eeq
where we estimate the gravitino density $\rho(t)$ at time $t$ by scaling back the known density 
at the equilibrium time $t_{eq}$ (with $8\pi G\rho_{rad}=8\pi G\rho_{mat} \sim 4\cdot 10^{-25}$ s$^{-2}$),
with the further assumption that at $t=t_{eq}$, most of the matter consisted of supermassive  
gravitinos, in line with our previous dark matter proposal \cite{MN1}. For this estimate we
also need to keep in mind that matter density scales as $a(t)^{-3}$ also during the radiation era
(while the radiation density scales as $a(t)^{-4}$). 

Gravitational collapse is expected to occur if the total energy is negative:
\beq\label{N}
\langle E_{kin} (t) \rangle \,+ \, \langle E_{pot} (t) \rangle \, < \, 0 \quad
\Ra \quad N \,>\,  \frac{T_{eq}\cdot 10^2\,{\rm m}}{G M_g^2} \,\sim \,  10^{12}
\eeq 
Importantly, the time $t$ drops out of this relation
because the temperature and the inverse average distance decrease in the same 
manner as a function of $t$ during the radiation era.  Let us stress that this
is only a very rough estimate: if the bound state is meta-stable, the collapse
can be delayed in such a way that a larger number of (anti-)gravitinos can be 
accrued. With (\ref{N}) the mass of the resulting mini-black hole comes out to be
\beq\label{Mlump}
m_{initial} \,\sim\, 10^{12} M_g \,\sim \, 10^{12} \MBPS \, \sim \, 10^3\, {\rm kg}
\eeq
By formula (\ref{evap}) the Hawking evaporation time for a black hole of this mass would be
\beq\label{evap1}
t_{evap} (m_{initial}) \,\sim \, 10^{-7}\, {\rm s}
\eeq
However, it is important now that Hawking evaporation is not the only process that
must be taken into account. There is a competing process which can in fact stabilize
the mini-black holes and their further evolution: it is the presence of the dense and hot plasma 
surrounding the black hole that can feed the growth of small black holes. More precisely, 
Hawking evaporation competes with accretion according to the following equation:
\beq\label{Mdot}
\frac{dm(t)}{dt}  \,=\,  C_0 G^2 \rho_{rad}(t)\cdot m^2(t)  
 \, -\, C_1 \frac{\MPL^3}{\tPL}\cdot \frac1{m^2(t)}
\eeq
where $C_0$ and $C_1$ are constants of $\cO(1)$. The first term on the r.h.s.
originates from the flux of the infalling radiation from the surrounding plasma,
which is $\propto 4\pi R^2(t) \rho_{rad}(t) c$ (with $c=1$) for a (time dependent) black hole of
radius $R(t) =2 Gm(t)$ (a `fudge factor' $C_0 =\cO(4\pi)$ can be included to account for 
the fact that not all the surrounding radiation falls in radially, but this is not essential  for our
argument).  The second term in (\ref{Mdot}) governs Hawking evaporation. 
For Hawking evaporation taking place in empty space we can ignore the first term 
on the r.h.s. of (\ref{Mdot}), and formula (\ref{evap}) follows directly. In that case
any microscopic black hole would disappear, and not be able 
to grow into a macroscopic black hole. The crucial difference with this standard 
scenario is embodied in the first term on the r.h.s. of (\ref{Mdot}) (which is usually 
disregarded in discussions of Hawking evaporation). This term takes into account the 
fact that the decay takes place in an extremely  hot surrounding plasma
whose density varies with time as $8\pi G\rho_{rad}(t) = 3/4t^2$. At the initially
extremely high temperatures of the radiation era the accretion can thus out-compete 
Hawking evaporation {\em even for very small black holes}. In terms of temperature
with the break-even point at $T_{rad} = T_{Hawking}$ where the radiation 
temperature $T_{rad}(t)$ at time $t$ can be read off from (\ref{Ekin}). The simple criterion for 
black hole accretion to overcome the rate for Hawking radiation reads 
\beq\label{Trad}
T_{rad} \,>\,  T_{Hawking} 
\eeq
This inequality is easy to achieve in the initially very 
dense and hot plasma where $T_{rad} \sim 10^{17} \GeV$. 
Later, it is a delicate issue because from (\ref{Mdot}) it follows that $m(t)$ can run away in either 
direction. This can also be directly seen by setting to zero the r.h.s. 
of (\ref{Mdot}): at time $t$ the break-even point occurs for
\beq\label{even}
m_0^4(t) \,\sim \, \frac{\MPL^3}{\tPL} \cdot \frac1{G^2 \rho_{rad}(t)} \,\propto \, t^2
\eeq
where we have used $\rho_{rad} = \frac{3}{32\pi G} t^{-2}$. Hence,
a mini-black hole of initial size $m_{initial}(t) > m_0(t)$ will be able to survive
and can start growing, whereas those of smaller mass decay.
Consequently, the earlier the bound state is formed, 
the smaller its initial mass can be. From these considerations and the time-independent
estimate (\ref{Mlump})  we can also derive a rough upper bound on the formation time, 
after which the radiation temperature is too low to stabilize mini-black holes against
Hawking evaporation. The maximal time $t_{max}$ is found by equating 
$m_0(t_{max}) \sim 10^3\,{\rm kg}$ from (\ref{Mlump}), which  yields the value
\beq\label{tmax}
     t_{max} = 10^{-20}\,{\rm s}
\eeq
Mini-black holes formed after this time can be expected to decay by Hawking radiation 
because $T_{rad}(t)  < T_{Hawking}$ for $t > t_{max}$. In summary,  the usual argument that 
small black holes would quickly decay via (\ref{evap}) no longer applies as long as
the inequality (\ref{Trad}) is obeyed.

Note that we invoke the `empirical' formula (\ref{Mdot}) mainly to argue that mini-black 
holes can form in such a way as to remain stable against Hawking evaporation at early times. 
In fact, this reasoning can be made more quantitative by substituting 
$\rho_{rad} = \frac{3}{32\pi G} t^{-2}$ into (\ref{Mdot}) which turns this equation into 
a simple differential equation that can be studied numerically.
However, because this formula is only approximate, and once the stability of the mini-black hole 
is ensured, we can switch to a classical description by means of an exact solution of Einstein's 
equations describing a Schwarzschild black hole in a radiatively expanding universe, to describe
its further evolution. This will be explained in the next section.

\section{Growth of black holes in radiation dominated universe}

Having motivated the assumption that small black holes stable against Hawking evaporation 
have formed in sufficient numbers early in the radiation dominated era we can proceed 
to study their evolution in this background. For this purpose we employ an 
{\em exact} solution of the Einstein equations, rather than the `phenomenological' 
formula (\ref{Mdot}). This solution can be regarded as a variant of the so-called McVittie 
solution \cite{McV0}; for more recent literature, see {\em e.g.} \cite{McV1,McV2,McV3,McV4,McV5,McV6} 
and references therein. The solution that we require here is conveniently presented 
in terms of conformal coordinates, by starting from the general ansatz 
\beq\label{metric0}
\rd s^2\,=\, a(\eta)^2\left[- C(\eta,r)\rd \eta^2+ \frac{\rd r^2}{C(\eta,r)}+r^2\rd\Omega^2\right]
\eeq
where $\eta$ is conformal time, which we use from now on as the time coordinate. 
$a(\eta)$ is the scale factor and $C=C(\eta,r)$ some function  to be specified. 
We will discuss the equations for the general ansatz elsewhere, but for the present
purposes it is enough to restrict to the special case, where $C$ depends only 
on the radial coordinate, {\em i.e.}  $C(\eta,r) \equiv C(r)$. Furthermore, since we are
here mainly interested in perfect fluids, for which $a(t) \sim t^{2/3(w+1)} \sim \eta^{2/(3w+1)}$, 
and more specifically, a radiation dominated universe, we right away specialize the scale 
factor to be 
\beq
a(\eta)=A\eta\quad \Longleftrightarrow \quad  t=\frac12 A \eta^2 \, .
\eeq 
where in our Universe $A \sim 4\cdot 10^{-5}\,$s$^{-1}$ (while $a(\eta)$ is dimensionless).
With these assumptions it is straightforward to compute the non-vanishing components of  
the Einstein tensor, and hence the components of the energy-momentum tensor, with the result
\bea\label{Tmn}
8\pi G \,T_{tt}(\eta,r) \,&=& \, - \frac1{\eta^2 r^2} \Big( C(r)C'(r) r\eta^2 + C^2(r)  \eta^2 - C(r) \eta^2 - 3r^2 \Big) \nn\\[2mm]
8\pi G\, T_{tr}(\eta,r) \,&=&\, \frac{C'(r)}{\eta C(r)} \nn\\[2mm]
8\pi G\,  T_{rr}(\eta,r) \,&=&\, \frac1{C(r)^2 r^2\eta^2} \Big( C(r) C'(r) r\eta^2 + C(r)^2 \eta^2 - C(r)\eta^2 + r^2\Big)  \nn\\[2mm]
8\pi G\,  T_{\theta\theta}(\eta,r) \,&=&\, \frac1{2C(r)\eta^2} \Big( C(r) C''(r) r^2 \eta^2 + 2C(r) C'(r) r\eta^2 + 2r^2 \Big) \nn\\[2mm] 
8 \pi G\, T_{\vp\vp}(\eta,r) \,&=&\, 8\pi G\, \sin^2\!\theta \, T_{\theta\theta}(\eta,r)
\eea
where, of course, $C'(r) \equiv dC(r)/dr$, {\em etc.} At this point, this is just an identity 
(the so-called `Synge trick' \cite{George}); in fact, such solutions trivially exist for {\em any} 
profile of the scale factor $a(\eta)$. The non-trivial part of the exercise is therefore
in ascertaining that the energy-momentum tensor resulting from this calculation does 
make sense {\em physically}. The requisite condition for a radiation dominated universe,
stated in the most general and coordinate independent way, is the vanishing of the 
trace of the energy-momentum tensor, {\em viz.}
\beq\label{Tmm}
T^\mu{}_\mu(\eta,r)  \,=\, \frac1{A^2\eta^2 r^2} \left[ \frac{\rd^2}{\rd r^2} \big( r^2 C(r) \big) - 2 \right] \, \stackrel{!}{=}\,  0
\eeq
This condition is solved by
\beq\label{Cr}
C(r) \,=\, 1 - \frac{2Gm}{r} + \frac{{G\cQ}^2}{r^2}
\eeq
with two integration constants $m$ (mass) and $\cQ$ (charge). Remarkably, the
metric (\ref{metric0}) comes out to be conformal to the Reissner-Nordstr\"om
metric not as a result of imposing the Einstein equations with an electromagnetic 
point charge source, but with the weaker and more general conformality 
constraint (\ref{Tmm})!  Taking $\cQ =0$ for simplicity (and also because we do not 
expect these black holes to carry significant amounts of electrical charge), the resulting 
solution describes the exterior region ($r>2Gm$) of a Schwarzschild black hole in a  
radiation dominated universe. We emphasize that there is absolutely no issue with the 
causal structure of this solution, because the conformal equivalence ensures that (for $\eta >0$) 
the global structure of the space-time outside the would-be horizon  $r=2Gm$ is the same as for the 
Schwarzschild solution, and the tracelessness of the energy momentum tensor holds right up 
to the would-be horizon (the black hole interpretation is also supported by the 
arguments in \cite{McV4,McV5}). However, there are some subtleties 
(apart from issues related to de Sitter space and cosmological horizons 
discussed in \cite{McV1,McV2,McV3,McV4,McV5}, which are
of no concern here) which have to do with the structure of the energy-momentum
tensor. Namely, as we show in the following section, closer inspection 
reveals the existence of an apparent `superluminal barrier' surrounding the surface $r=2Gm$,
and shielding the would-be horizon from the outside observer. 

For the physical mass of the black hole we take the formula
\beq\label{mass}
\frac1{2\pi} \int \rd\theta a(\eta) r \Big|_{r=2Gm}\,=\, 2 Gm a(\eta)\quad \Ra 
\quad m(\eta)=m a(\eta)
\eeq
keeping in mind that the observer at infinity will in addition measure the integrated
matter density outside the apparent horizon, so the above formula is really
a lower bound on the total mass accretion. The total mass therefore grows (at least) 
linearly with the scale factor, and this is also consistent with the fact that $T_{tr} \neq 0$.
The formula (\ref{mass}) gives (with $\eta = \eta_{initial}$)
\beq
m=\frac{m_{initial}}{a_{initial}}
\eeq
where $m_{initial}$ is any value compatible with the 
lower bound following from (\ref{even}), 
and $a_{initial}$ is the scale factor at the time when the black hole forms.
The mass accretion described by (\ref{mass}) is also evident from the 
non-vanishing mixed component $T_{tr}$ in (\ref{Tmn}) which states that 
there is energy flow into the black hole from the surrounding radiative medium.
During the radiation era there is, in fact, an unlimited supply of `food' for the black hole
to swallow. This supply will dry up only when inhomogeneities are formed, after
which the accretion works in the more standard form. 

Evolving the initial mass (\ref{Mgrav}) with the formula (\ref{mass}) we calculate the
final mass at the equilibrium time (assuming that $\eta_{initial}\sim \eta_{\rm Pl}$)
\beq\label{mfinal}
m_{final} \,\sim \, 
m_{initial} \left(\frac{\eta_{eq}}{\eta_{\rm Pl}} \right) \,\sim\,  10^{30}\ {\rm kg}
\,\sim\, M_{\odot}
\eeq
with $\eta_{eq} \sim 2\cdot 10^8\,$s and $\eta_{\rm Pl} \sim 10^{-19}\,$s.
This estimate applies to mini-black holes formed very early in the radiation era (for $\eta\ll \eta_{max}$).
The same calculation for a mini-black hole at the latest possible time $\eta_{max}$ given by (\ref{tmax}) 
also yields a lower bound for the final mass of the primordial black hole upon exit from the radiation era,  
\beq\label{Mfinal}
\sqrt{m_{initial} \MPL} \,<\,  m_{final} \,<\,  \MPL \left(\frac{\eta_{eq}}{\eta_{\rm Pl}}\right)
\eeq
or
\beq
10^{11}\,{\rm kg} \,<\, m_{final} \, < \, M_\odot
\eeq
This inequality restricts the possible mass range for primordial black holes at the equilibrium time.

The above analysis can be repeated for matter dominated and exponentially 
expanding universes, respectively. In this case we need the angular Killing vectors 
$k^\mu_\theta\pa_\mu $ and $k^\mu_\vp\pa_\mu$ to state the pertinent conditions 
in a generally covariant way. In the matter dominated era we have 
\beq
a(\eta)=B^2\eta^2\;\; ,\quad
 T_{\mu\nu}k_\th^\mu k_\th^\nu\,=\,  T_{\mu\nu}k_\phi^\mu k_\phi^\nu\,=\,0\quad
  \Ra\; C(r)\,=\,1-\frac{2Gm}{r}
\eeq
where we utilize the Killing vectors to state the condition of vanishing pressure.
Because this solution does not allow for a non-vanishing charge, this provides another 
reason for setting $\cQ = 0$ in (\ref{Cr}), in order to allow for a smooth transition from the 
radiation dominated to the matter dominated phase.
From this we see that the primordial black holes will
continue to grow with the scale factor also in the early part of the matter dominated 
phase, absorbing radiation {\em and} matter, as long as there are no significant 
inhomogeneities. After the distribution of
matter develops inhomogeneities, the further evolution of black holes   proceeds in
the standard fashion. In other words, the range of mass values in (\ref{Mfinal})
corresponding to time $t= t_{eq}$, only represent a lower limit, as the black holes will 
continue to accrete mass in significant amounts until inhomogeneities start forming.

Finally, for an exponentially expanding universe we have
\beq
a(\eta)=\frac{1}{H(\eta_\infty - \eta)} \;\;, \quad  
T^\mu{}_\mu \,=\, 2\big(T_{\mu\nu}k_\th^\mu k_\th^\nu+T_{\mu\nu}k_\phi^\mu k_\phi^\nu \big) \quad
 \Ra\;  C(r)\,=\, 1-\frac{2Gm}{r}-C_Hr^2
\eeq
Please note that for $C_H \neq 0$ this is {\em not} the well known Kottler solution (that is,
de Sitter space in static coordinates). We stress again that for $C(\eta,r) = C(r)$ and with
$\cQ=0$ and $C_H = 0$ the causal structure of the space-time is the same as for an ordinary 
black hole space-time, and only in this case we can have a smooth transition between all phases.

\section{Energy-momentum tensor}

To gain further insight into the physical properties of our solution let us examine
the energy-momentum tensor (\ref{Tmn}) a bit more closely. Following \cite{Weinberg}
we parametrize the latter as
\bea\label{Tmn1}
T_{\mu\nu} \,&=&\,  p g_{\mu\nu} + (p+\rho) u_\mu u_\nu -
    \Pi_{\mu\rho} Q^\rho u_\nu   - \Pi_{\nu\rho} Q^\rho u_\mu     \nn\\[2mm]
     &&  - \, \zeta_1 \, \Pi_\mu{}^\rho \Pi_\nu{}^\sigma 
     \left (\nabla_\rho u_\sigma + \nabla_\sigma u_\rho - 
         \frac23 g_{\rho\sigma} \nabla^\lambda u_\lambda \right)
      \, - \, \zeta_2 \, \Pi_{\mu\nu} \nabla^\lambda u_\lambda 
\eea
where $u^\mu u_\mu = -1$, $Q^\mu$ is the heat flow, and $\zeta_1$ and $\zeta_2$ are the 
shear and bulk viscosity, respectively. All variables are assumed to depend on $\eta$ 
and $r$ only. The projector is defined by
\beq
\Pi_{\mu\nu} = g_{\mu\nu} + u_\mu u_\nu
\eeq
We will now match the energy-momentum tensor (\ref{Tmn}) to this formula. For
simplicity we assume
\beq\label{visc}
\zeta_1 = \zeta_2 = 0
\eeq
We also write $q_\mu \equiv \Pi_{\mu\nu} Q^\nu$ (so that $u^\mu q_\mu = 0$), so that
the energy momentum tensor simplifies to
\beq\label{Tmn3}
T_{\mu\nu} \,=\,  p g_{\mu\nu} + (p+\rho) u_\mu u_\nu - q_\mu u_\nu   -  q_\nu u_\mu
\eeq
The assumption of vanishing viscosity coefficients (\ref{visc}) is certainly justified after 
baryogenesis (that is $t > 10^{-12}\,$s), when the number of photons by far exceeds 
the number of other particles in the plasma (for instance, $n_\gamma \sim 10^{10}\, n_{b}$).
While the condition (\ref{Tmm}) leaves $\zeta_1$ undetermined, we could in principle 
also admit a non-vanishing $\zeta_2 \neq 0$, that is, self-interacting conformal 
matter ({\em e.g.} self-interacting massless scalar fields). In that case the relation 
$\rho=3p$ derived below would no longer hold even with vanishing $T^\mu{}_\mu$.

For the comparison we write out (\ref{Tmn}) explicitly for the solution (\ref{Cr}) 
(with $\cQ=0$), which gives
\bea\label{Tmn2}
8\pi G \, T_{tt} \,&=&\, \frac{3\,\dot a^2}{a^2} \,=\, \frac3{\eta^2}  \nn\\[2mm]
8\pi G \, T_{rr} \,&=&\, \frac{r^2( -2a \ddot a + \dot a^2)}{a^2(r-2Gm)^2} 
      \,=\, \frac{r^2}{\eta^2(r- 2Gm)^2}  \nn\\[2mm]
8 \pi G \, T_{rt} \,&=&\, \frac{2Gm \dot a}{ar(r-2Gm)} 
    \,=\, \frac{2Gm}{r\eta(r-2Gm)}  \nn\\[2mm]
T_{\theta\theta} \,&=&\, r^2 T_{rr} \quad , \quad T_{\vp\vp} = r^2 \sin^2\!\theta\,  T_{rr}
\eea
Comparing (\ref{Tmn2}) and (\ref{Tmn3}) we read off the unknown quantities on the r.h.s.
of (\ref{Tmn3}); we find
\bea\label{uq}
u_\mu(\eta,r) \,&=&\, A\eta
\left(\sqrt{\frac{r - 2Gm}{r}} \cosh\xi\,,\, \sqrt{\frac{r}{r -2Gm}}\sinh\xi\,,\,0\,,\,0\right) \nn\\[2mm]
q_\mu (\eta,r)\,&=&\, A\eta q(\eta,r) 
\left( \sqrt{\frac{r - 2Gm}{r}}\sinh\xi \,,\, \sqrt{\frac{r}{r -2Gm}}\cosh\xi\,,\,0\,,\,0\right) 
\eea
where
\beq
\tanh \xi \,=\, \frac{Gm\eta}{r^2}     \qquad\quad  (\Rightarrow \; \xi>0)
\eeq
and
\beq\label{q}
q(\eta,r) \,=\, 2p(\eta,r)  \tanh\xi 
\eeq
(with $m \equiv m_{initial}$). The density and pressure are given by 
\beq\label{rho}
\rho(\eta,r) = 3 p(\eta,r) \qquad \mbox{with} \quad p(\eta,r) \,=\, \frac{r}{A^2 \eta^2 (r- 2Gm)} 
\eeq
as expected for a radiation dominated universe. We stress that there are 
no pathologies here of the kind encountered in some of the previous literature 
on McVittie-type solutions. In particular, the energy density $\rho(\eta,r)$ is strictly positive
for $r> 2Gm$ and at all times $\eta > 0$. Moreover, because $q$ is positive from (\ref{q}), the
radial component of $q^\mu$ in (\ref{uq}) is also positive, which means that the radial
heat flow is {\em inward directed}, explaining why the mass of the black hole {\em grows} 
with time.

To keep $\xi$ real we must demand
\beq
\tanh \xi = \frac{Gm\eta}{r^2} \,< \,1 \quad \Rightarrow \quad r\, >\, \sqrt{ Gm\eta} \quad (> 2Gm )
\eeq
For $r^2 \rightarrow Gm\eta$ the average velocity of the infalling matter reaches the speed of 
light, and the expansion (\ref{Tmn1}) in powers of $u_\mu$ and its derivatives breaks 
down. Consequently, while the solution (\ref{metric0}) remains valid down to
$r=2Gm$, the expressions (\ref{uq}), (\ref{q}) and (\ref{rho}) become meaningless 
in the region $2Gm < r < \sqrt{Gm\eta}\,$ because of apparently superluminal propagation 
(similar conclusions regarding superluminality were already reached in \cite{McV1}). 
Likewise the components of the heat flow $q^\mu$ diverge for $\tanh\xi\rightarrow 1$,
indicating an apparent divergence of the temperature in this limit. This is also an
unphysical feature in view of the breakdown of the expansion (\ref{Tmn1}).
Physically it is tempting to interpret this result as implying that the would-be horizon 
is shielded from the outside observer by a `blanket' at $r=\sqrt{Gm\eta}$, whose 
extension grows with cosmic time $\eta$. However, in recent work \cite{HS} it is
argued that the gradient expansion (\ref{Tmn1}) must be replaced by a different expansion;
adapting these arguments to the present case we conclude that the solution 
can, in fact, remain meaningful all the way down to $r=2Gm$. Because of the breakdown of the 
expansion (\ref{Tmn1}), also the apparent `firewall' ($\equiv$ divergent energy density $\rho$) 
on the would-be horizon $r=2Gm$ is an unphysical feature  
(we have checked that by re-instating the $\eta$-dependence
in the metric coefficient $C(\eta,r)$ and setting up an appropriate 
expansion near the would-be horizon one can eliminate this divergence).
This is just as well, because
otherwise the total mass at infinity (which includes the integrated energy density for
$r>2Gm$) would diverge, as $\rho(\eta,r)$ has a non-integrable singularity at $r = 2Gm$.
At any rate these arguments show that the actual mass value for the black hole will
exceed the estimated value (\ref{mfinal}) if the matter contributions outside the
horizon are taken into account, thus further enhancing the growth of primordial black holes.

{\section{Conclusions}
In this paper we have proposed a new mechanism to explain the emergence of supermassive
primordial black holes during the radiation period. The key element here is the conjectured
existence of very massive particles stable against decay into Standard Model matter,
that can `condense' into bound states sufficiently early in the radiation period
which can subsequently collapse to black holes. Our proposal is chiefly motivated by the 
possible explanation of the observed spectrum of 48 spin-$\frac12$ fermions in the
Standard Model that was put forward in our previous work \cite{MN2,KN,MN0}, and is thus
subject to independent falsification if {\em any} new fundamental spin-$\frac12$ fermions were 
to show up in future collider searches. In addition, we have derived a new solution of 
Einstein's equations describing the growth of black holes in a dense and hot plasma 
through inflow of radiation. This exact solution could be useful also in other contexts.

\vspace{1cm} 
\noindent
 {\bf Acknowledgments:} We would like to thank B.F. Schutz for correspondence and helpful
comments, and the referee for suggesting several  improvements in the original version.
K.A.M. thanks AEI for hospitality and support; he was 
 partially supported by the Polish National Science Center grant DEC-2017/25/B/ST2/00165.
 The work of  H.N. has received funding from the European Research 
 Council (ERC) under the  European Union's Horizon 2020 research and 
 innovation programme (grant agreement No 740209).

\end{document}